\documentclass[twocolumn,trackchanges]{aastex631}

\usepackage{newtxtext,newtxmath}

\newcommand{\mosel}{{\tt MOSEL}}

\newcommand{\zfourge}{{\tt ZFOURGE}}
\newcommand{\jades}{{\tt JADES}}

\newcommand{\oiii}{[\hbox{{\rm O}\kern 0.1em{\sc iii}}]\,5007}
\newcommand{\nii}{[\hbox{{\rm N}\kern 0.1em{\sc ii}}]\,6583}
\newcommand{\ciii}{[\hbox{{\rm C}\kern 0.1em{\sc iii}}]\,1907,1909}
\newcommand{\civ}{\hbox{{\rm C}\kern 0.1em{\sc iv}}\,1550}
\newcommand{\oiiiuv}{[\hbox{{\rm O}\kern 0.1em{\sc iii}}]\,1660,1666}
\newcommand{\oiiite}{[\hbox{{\rm O}\kern 0.1em{\sc iii}}]\,4363}
\newcommand{\heii}{\hbox{{\rm He}\kern 0.1em{\sc ii}}\,1640}
\newcommand{\hii}{\hbox{{\rm H}\kern 0.1em{\sc ii}}}

\newcommand{\hb}{\hbox{{\rm H}\kern 0.1em{\sc $\beta$}}}
\newcommand{\ha}{\hbox{{\rm H}\kern 0.1em{\sc $\alpha$}}}
\newcommand{\ciiradio}{[\hbox{{\rm C}\kern 0.1em{\sc ii}}]$_{\rm 158 \mu m}$}

\newcommand{\SiII}{[\hbox{{\rm Si}\kern 0.1em{\sc ii}}]}

\newcommand{\OI}{[\hbox{{\rm O}\kern 0.1em{\sc i}}]}

\newcommand{\oiiiradio}{[\hbox{{\rm O}\kern 0.1em{\sc iii}}]$_{\rm 88 \mu m}$}

\newcommand{\msun}{${\rm M_{\odot}}$}

\newcommand{\lya}{\hbox{{\rm Lyman-}\kern 0.1em{\sc $\alpha$}}}
\newcommand{\logmstar}{$\log(M_*/{\rm M}_{\odot})$}

\newcommand{\nEELGs}{76}
\newcommand{\nEELGsJades}{19}
\newcommand{\ncontrol}{1712}
\newcommand{\ncontrolJades}{275}

\begin{document}
	
	\title{MOSEL survey: JWST reveals major mergers/strong interactions drive the extreme emission lines in the early universe}

	\author[0000-0002-8984-3666]{Anshu Gupta}
	\affiliation{International Centre for Radio Astronomy Research (ICRAR), Curtin University, Bentley WA, Australia}
	\affiliation{ARC Centre of Excellence for All Sky Astrophysics in 3 Dimensions (ASTRO 3D), Australia}
	
	\author[0000-0003-2035-3850]{Ravi Jaiswar}
	\affiliation{International Centre for Radio Astronomy Research (ICRAR), Curtin University, Bentley WA, Australia}
	\affiliation{ARC Centre of Excellence for All Sky Astrophysics in 3 Dimensions (ASTRO 3D), Australia}

	\author[0000-0002-9495-0079]{Vicente Rodriguez-Gomez}
	\affiliation{Department of Physics and Astronomy, Johns Hopkins University, Baltimore, MD 21218, USA} 
	\affiliation{Instituto de Radioastronom\'ia y Astrof\'isica, Universidad Nacional Aut\'onoma de M\'exico, A.P. 72-3, 58089 Morelia, Mexico}
	
	\author[0000-0001-6003-0541]{Ben Forrest}
	\affiliation{Department of Physics and Astronomy, University of California Davis, One Shields Avenue, Davis, CA, 95616, USA}
	
	\author[0000-0001-9208-2143]{Kim-Vy Tran}
	\affiliation{School of Physics, University of New South Wales, Sydney, NSW 2052, Australia}
	\affiliation{ARC Centre of Excellence for All Sky Astrophysics in 3 Dimensions (ASTRO 3D), Australia}
	\affiliation{Center for Astrophysics $|$ Harvard \& Smithsonian, Cambridge, MA }
	
	\author[0000-0003-2804-0648]{Themiya Nanayakkara}
	\affiliation{Centre for Astrophysics and Supercomputing, Swinburne University of Technology, Hawthorn, VIC 3122, Australia}
	\affiliation{ARC Centre of Excellence for All Sky Astrophysics in 3 Dimensions (ASTRO 3D), Australia}
	
	\author[0000-0001-9414-6382]{Anishya Harshan}
	\affiliation{University of Ljubljana, Department of Mathematics and Physics, Jadranska ulica 19, SI-1000 Ljubljana, Slovenia}
	
	\author[0000-0001-9759-4797]{Elisabete da Cunha}
	\affiliation{International Centre for Radio Astronomy Research, University of Western Australia, 35 Stirling Hwy, Crawley, WA 6009, Australia}
	\affiliation{ARC Centre of Excellence for All Sky Astrophysics in 3 Dimensions (ASTRO 3D), Australia}

	\author[0000-0003-1362-9302]{Glenn G. Kacprzak}
	\affiliation{Swinburne University of Technology, Hawthorn, VIC 3122, Australia}
	\affiliation{ARC Centre of Excellence for All Sky Astrophysics in 3 Dimensions (ASTRO 3D), Australia}

	\author[0000-0002-3301-3321]{Michaela Hirschmann}
	\affiliation{Institute of Physics, GalSpec, Ecole Polytechnique Federale de Lausanne, Observatoire de Sauverny, Chemin Pegasi 51, 1290 Versoix, Switzerland} 
	\affiliation{INAF, Astronomical Observatory of Trieste, Via Tiepolo 11, 34131 Trieste, Italy}

	\begin{abstract}

		Extreme emission line galaxies (EELGs), where nebular emissions contribute 30-40\% of the flux in certain photometric bands, are ubiquitous in the early universe ($z>6$). We utilise deep NIRCam imaging from the JWST Advanced Deep Extragalactic Survey (\jades) to investigate the properties of companion galaxies (projected distance $<40\,kpc$,  $|dv|<10,000$\,km/s) around EELGs at $z\sim3$. Tests with TNG100 simulation reveal that nearly all galaxies at $z=3$ will merge with at least one companion galaxy selected using similar parameters by $z=0$. The median mass ratio of the most massive companion and the total mass ratio of all companions around EELGs is  more than 10 times higher the control sample. Even after comparing  with a stellar mass and stellar mass plus specific SFR-matched control sample, EELGs have three-to-five times higher mass ratios of the brightest companion and total mass ratio of all companions. Our measurements suggest that EELGs are more likely to be experiencing strong interactions or undergoing major mergers irrespective of their stellar mass or specific SFRs. We suspect that gas cooling induced by strong interactions and/or major mergers could be triggering the extreme emission lines, and the increased merger rate might be responsible for the over-abundance of EELGs at $z>6$.

	\end{abstract}
	
	
	\section{Introduction} \label{sec:intro}

	In the past decade, deep photometric surveys with {\it Spitzer} and {\it Hubble space telescope} (HST) revealed more than two orders increase in the \oiii+\hb\ equivalent width (EW) of galaxies between $z=0-6$ \citep{labbe_2013_SPECTRALENERGYDISTRIBUTIONS, roberts-borsani_2016_GALAXIESREDSPITZER, barro_2019_CANDELSSHARDSMultiwavelength, mainali_2019_RELICSSpectroscopyGravitationallylensed, endsley_2020_IiiEquivalentWidth, gupta_2022_MOSELSurveyExtremely}. Direct observations with the James Webb Space Telescope (JWST) in the past year have confirmed that   80\% of galaxies have \oiii+\hb\ EW $>800$\,\AA; almost three times the EW of a typical star-forming galaxy at $z\sim2$ \citep{endsley_2023_StarformingIonizingProperties, cameron_2023_JADESProbingInterstellar, tang_2023_JWSTNIRSpecSpectroscopy, rinaldi_2023_MIDISStrongHv}. Thus, understanding the physical origin of extreme emission lines is becoming increasingly important to understand the early galaxy evolution. 
	
	Extreme emission line galaxies (EELGs) at lower redshifts typically have low stellar masses and high star formation rates  \citep{atek_2011_VERYSTRONGEMISSIONLINE, wel_2011_EXTREMEEMISSIONLINEGALAXIES, maseda_2013_CONFIRMATIONSMALLDYNAMICAL, maseda_2014_NatureExtremeEmission, chevallard_2018_PhysicalPropertiesHionizingphoton, tang_2019_MMTMMIRSSpectroscopy, gupta_2022_MOSELSurveyExtremely, lumbreras-calle_2022_JPLUSUncoveringLarge} and might be undergoing first burst in their  star formation history \citep{cohn_2018_ZfourgeExtreme5007, endsley_2023_StarformingIonizingProperties}. \cite{reddy_2018_MOSDEFSurveySignificant} show that evolution of stellar mass and star formation main-sequence is sufficient to explain the moderate increase in the \oiii\ EW  between $z=0\ {\rm to}\ z\sim2$. Although, it is possible that increased stochasticity in the star formation history is sufficient to explain the overabundance of EELGs in the early universe \citep{endsley_2023_StarformingIonizingProperties, dressler_2023_BuildingFirstGalaxies}. 
	
	Mergers of gas-rich galaxies and galaxy-galaxy interactions can funnel gas into the galactic center, boosting the star formation \citep{sanders_1996_LuminousInfraredGalaxies, hayward_2013_SubmillimetreGalaxiesHierarchical, sparre_2017_UnorthodoxEvolutionMajor, moreno_2019_InteractingGalaxiesFIRE2}.  Cosmological zoom-in simulations show that scatter in star-forming main-sequence is due to mergers and gas accretion events \citep{tacchella_2016_ConfinementStarformingGalaxies, sparre_2017_StarBurstsFIRE, torrey_2018_SimilarStarFormation},  and the  gas accretion events also sets the intermediate-scale ($<1$ Gyrs) variability in the star formation history  \citep{sparre_2017_StarBurstsFIRE, tacchella_2020_StochasticModellingStarformation}. Mergers and galaxy-galaxy interaction can induce the circumgalactic medium gas to cool down, boosting the star formation rate by $30-40\%$ \citep{moreno_2019_InteractingGalaxiesFIRE2, sparre_2022_GasFlowsGalaxy}.
	
	Traditionally, asymmetry and smoothness in the stellar light profile of galaxies is used to identify galaxies undergoing major mergers \citep{conselice_2003_RelationshipStellarLight, lotz_2004_NewNonparametricApproach}. However, mock-imaging of simulated galaxies show that the accuracy of morphology indicators varies between $30-60\%$ and depends significantly on the choice of photometric filter \citep{rose_2023_IdentifyingGalaxyMergers}. Sophisticated photometric and/or deep spectroscopic observations have been used to identify pairs of galaxies that  would end up merging together at some point \citep{lin_2004_DEEP2GalaxyRedshift, watson_2019_GalaxyMergerFractions, duncan_2019_ObservationalConstraintsMergera}.

	In this letter, we use the deep NIRCam photometry from the \jades\ survey \citep{bunker_2023_JADESNIRSpecInitiala, hainline_2023_CosmosItsInfancy, rieke_2023_JADESInitialData, eisenstein_2023_OverviewJWSTAdvanceda} to analyse the properties of companions around EELGs at $z\sim3$. The EELGs were identified in the \zfourge\ survey \citep{straatman_2016_FourStarGalaxyEvolution} using composite spectral energy distribution fitting \citep{forrest_2018_ZFOURGEUsingComposite} and later confirmed as part of the Multi-Object Spectroscopic of Emission Line (\mosel) survey  \citep{tran_2020_MOSELStrongOiii, gupta_2022_MOSELSurveyExtremely}. We find that EELGs are more likely to have similar stellar mass companions than the control sample, suggesting they are more likely to be either undergoing major mergers or experiencing strong interactions.

	\section{Data}\label{sec:data}

	The EELGs sample is selected from the FourStar Galaxy Evolution survey \citep[\zfourge;][]{straatman_2016_FourStarGalaxyEvolution},   which uses medium-band filters in J and H bands on the Fourstar instrument ($13' \times 13'$) on the Magellan telescope to reach a photometric redshift accuracy of $<2\%$ at $1.5<z<4$ \citep{nanayakkara_2016_ZFIREKECKMOSFIRE, tran_2020_MOSELStrongOiii}.
	\cite{forrest_2018_ZFOURGEUsingComposite} identified \nEELGs\ EELGs between $2.5<z<4$  in the Chandra Deep field South (CDFS) using composite spectral energy distribution fitting. For comparison, we also  select \ncontrol\ galaxies within the same redshift range and K-band signal-to-noise (S/N) $>10$ from the \zfourge\ survey.

	The first data release (DR1) of \jades\ presents unprecedented depth (4.5nJy at $5\sigma$ in F444W) for $\sim 25\,{\rm arcminute}^2$ 
	\citep{bunker_2023_JADESNIRSpecInitiala, hainline_2023_CosmosItsInfancy, rieke_2023_JADESInitialData, eisenstein_2023_OverviewJWSTAdvanceda} within the CDFS (MAST:\dataset[10.17909/8tdj-8n28]{http://dx.doi.org/10.17909/8tdj-8n28},\dataset[10.17909/fsc4-dt61]{http://dx.doi.org/10.17909/fsc4-dt61}). We match the brightest galaxy in the F444W filter within $0.5''$ to cross match \zfourge\ and \jades\ samples, to account for small astrometric woffset ($\sim 0.3''$) between the two surveys.   We find only \nEELGsJades\ out of \nEELGs\ EELGs and \ncontrolJades\ out of \ncontrol\ control galaxies in the JADES DR1 because of its relatively smaller footprint compared to the \zfourge\ survey. 
	
	\begin{figure}
		\centering
		\tiny
		\includegraphics[scale=0.55, trim=0.0cm 0.0cm 0.0cm 0.0cm,clip=true]{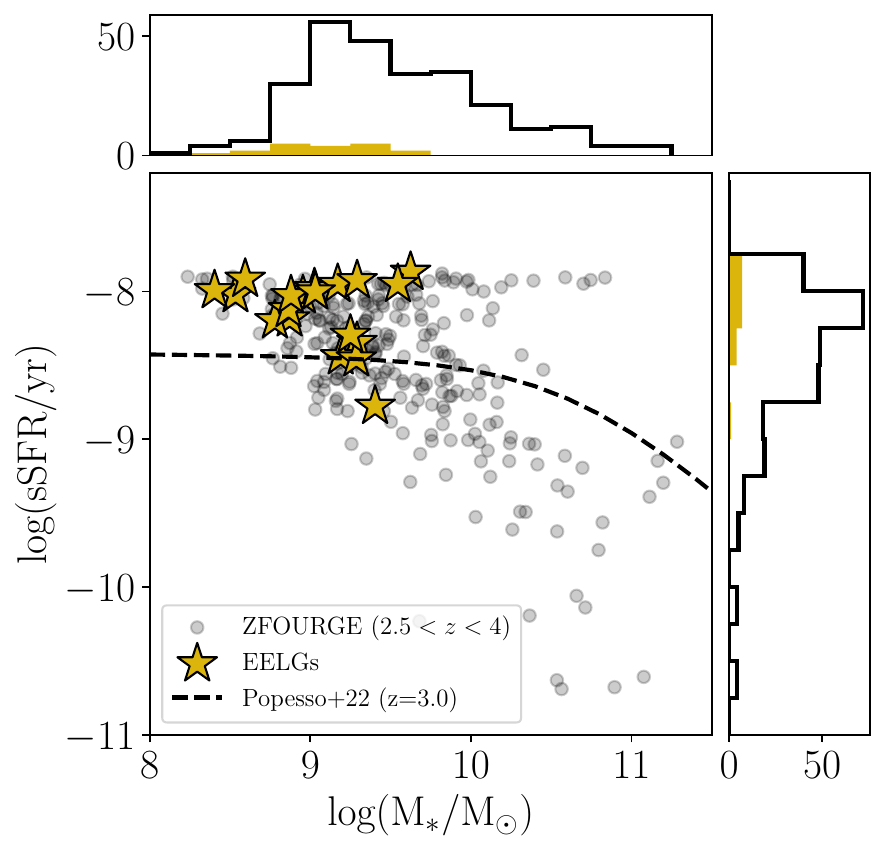}		
		\caption{Stellar mass and  specific SFRs distribution of EELGs (golden stars) and full control sample (grey circles). The SFR are derived from MAGPHYS SED fitting code. The dashed line correspond to the star forming main sequence at $z=3$ by \citep{popesso_2022_MainSequenceStarforming}. At a fixed stellar mass EELGs have 0.3\, dex higher sSFR compared to the control galaxies.}
		\label{fig:companions_prop}
	\end{figure}
	
	\section{Analysis}
	
	\subsection{Photometric redshifts}\label{sec:redshift}
	To identify companion galaxies, we need redshifts for EELGs, the control sample and their possible companions.  Only 12/19 [66/275] of EELGs [control]  have  spectroscopic redshifts either from the  \mosel\ survey  \citep{tran_2020_MOSELStrongOiii, gupta_2022_MOSELSurveyExtremely} or the \jades\ DR1,  which has collected spectroscopic redshifts from many surveys in the literature plus the FRESCO survey \citep{oesch_2023_JWSTFRESCOSurvey} and NIRSpec observation by JADES \citep{rieke_2023_JADESInitialData}. 
	
	\begin{figure*}
		\centering
		\tiny
		\includegraphics[scale=0.31, trim=0.2cm 0.2cm 0.2cm 0.0cm,clip=true]{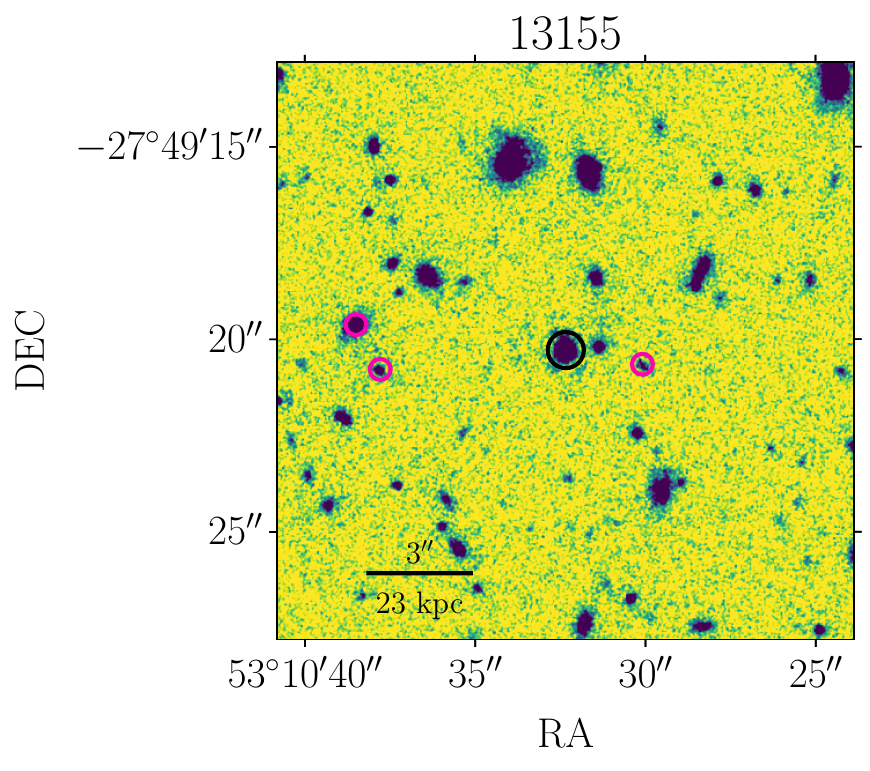}
		\includegraphics[scale=0.31, trim=1cm 0.2cm 0.2cm 0.0cm,clip=true]{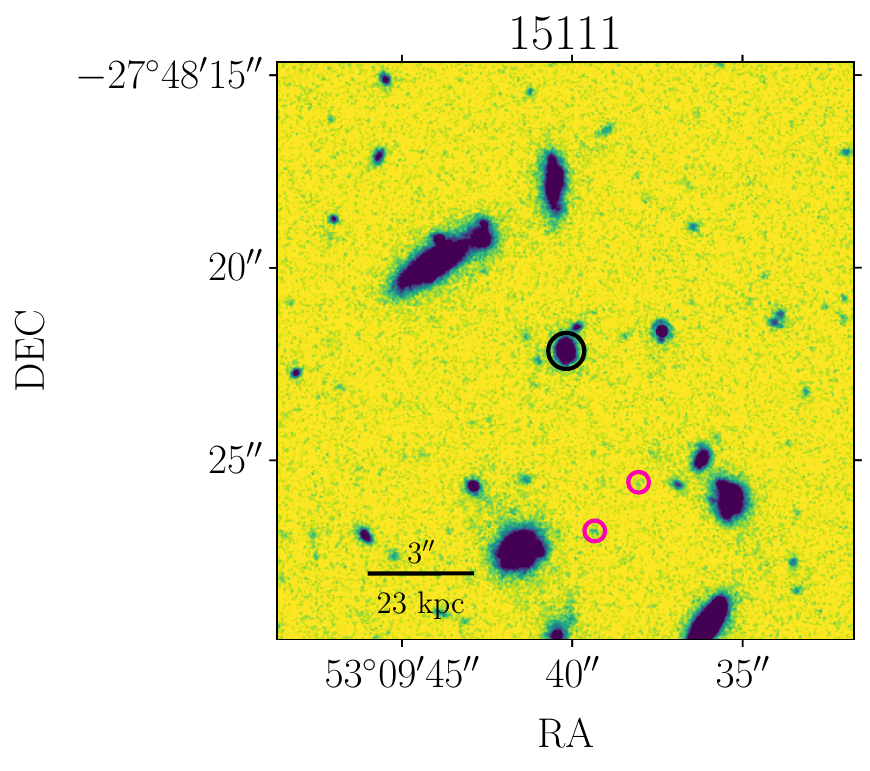}	
		\includegraphics[scale=0.31, trim=1cm 0.2cm 0.2cm 0.0cm,clip=true]{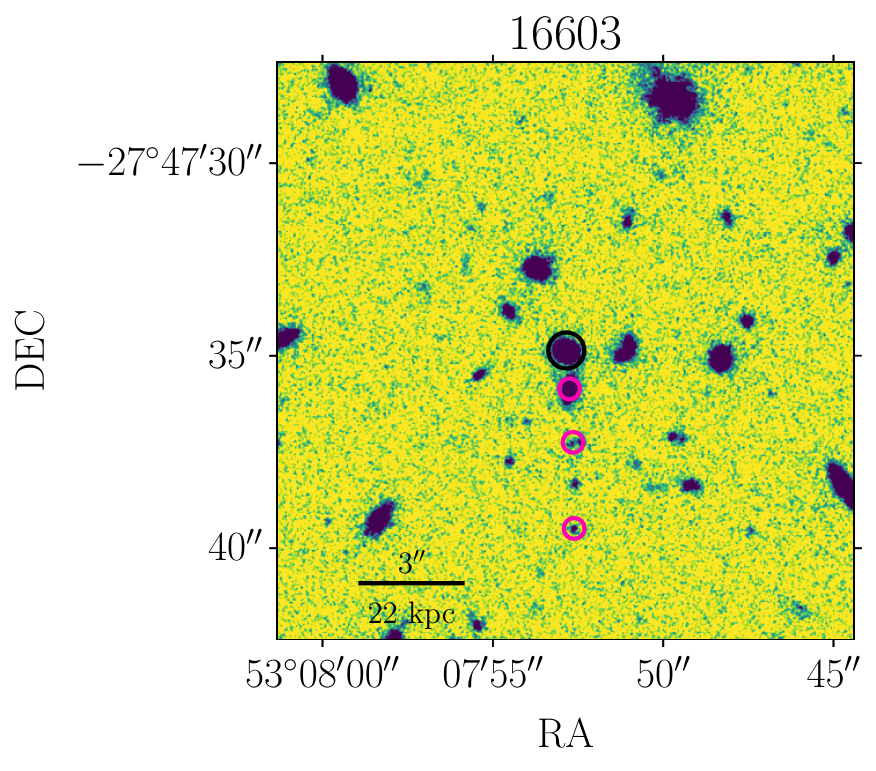}	
		\includegraphics[scale=0.31, trim=1cm 0.2cm 0.2cm 0.0cm,clip=true]{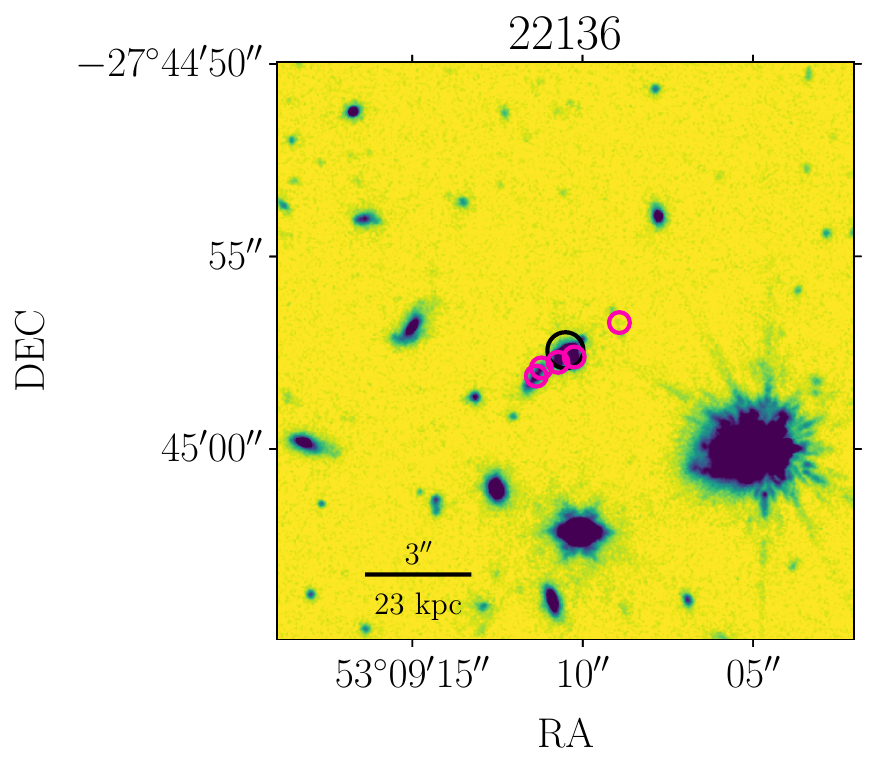}	
		\caption{Example $JWST$/NIRCam F444W images from the \jades\ survey \citep{rieke_2023_JADESInitialData} for four EELGs (black circles and  their companions (pink circles) showing the variety of companions around the target galaxy.}
		\label{fig:2d_maps}
	\end{figure*}
	
	Both \zfourge\ and \jades\ survey use EAZY \citep{brammer_2008_EAZYFastPublic} to calculate photometric redshifts. Additional stellar templates with younger stellar ages  were added by the \jades\ team \citep{hainline_2023_CosmosItsInfancy}. We do not find any systematic difference in the  photometric redshifts from \jades\ and \zfourge\ surveys.  The average offset between the photometric redshift and spectroscopic redshift for EELGs [control] is $<z_{\rm spec}-z_{\rm phot}>=-0.09$ [-0.01], $\sigma_{NMAD} = 0.017$ [0.019] and zero outliers [3\%] from the \jades\ survey. For consistency, we use photometric redshifts from the \jades\ DR1 throughout this letter when spectroscopic redshifts are unavailable.

	\subsection{Spectral energy distribution}\label{sec:sed}
	We use the MAGPHYS \citep{dacunha_2008_SimpleModelInterpret, dacunha_2015_ALMASurveySubmillimeter} spectral energy distribution (SED) fitting code with BC03 stellar population synthesis model \citep{bruzual_2003_StellarPopulationSynthesis}, delayed exponentially declining star formation history model,  \cite{charlot_2000_SimpleModelAbsorptiona} dust attenuation law  to derive physical properties.  The results presented in this letter were derived only using the 23-band photometry from the JADES DR1 (HST + broad-band  \jades\ + JWST Extragalactic Medium-band Survey \citep[JEMS,][]{williams_2023_JEMSDeepMediumband})  and fixing the redshift to best redshift determined in Section \ref{sec:redshift}  because most companions are undetected in the \zfourge\ survey.  
	
	Figure \ref{fig:companions_prop} shows the stellar mass and specific star formation rate (sSFR) distribution of our primary targets.  As expected EELGs have higher sSFRs and lower stellar mass  than the control sample by about 0.3\,dex \citep[Table \ref{tb:companion_properties},][]{forrest_2018_ZFOURGEUsingComposite, gupta_2022_MOSELSurveyExtremely}. MAGPHYS does not include emission lines, exclusion of which can increase the estimated stellar masses by up to 0.5\,dex especially for EELGs \citep{forrest_2018_ZFOURGEUsingComposite}.  However, our stellar masses do not change significantly ($<0.04$\,dex) after removing filters (F277W or F277W plus F356W) that will be contaminated with \oiii\ and \ha\ emission lines. We suspect inclusion of longer wavelength filters  (F356W and F444W) and their broadness minimise the effect of emission lines on stellar mass estimates.

	\subsection{Companion galaxies}\label{sec:companions}
	We use the distribution of galaxies in phase-space to identify companions around the target galaxies. Spectroscopic studies of pair fractions measurements typically employ projected distances of 20 to 50 kpc and velocity offsets of $|dv| < 500$\,km/s to identify interacting pairs of galaxies \citep{patton_2000_NewTechniquesRelating,  mantha_2018_MajorMergingHistory}. Studies relying on photometric redshifts adopt a more statistical approach to account for the uncertainty in photometric redshift measurements \citep{lopez-sanjuan_2015_ALHAMBRASurveyAccurate, duncan_2019_ObservationalConstraintsMergera, watson_2019_GalaxyMergerFractions}. 
	
	The companion analysis is restricted to all galaxies detected at S/N$>5$ in F444W filter. The choice of F444W filter ensures that companion galaxies are well detected in all shorter wavelength filters, which is necessary for accurate spectral energy distribution modelling and photometric redshift estimation. Only 4\% of the galaxies with S/N$>5$ in F444W filter have spectroscopic redshifts and less than 10\% have counterpart in the \zfourge\ survey. Thus, we use the photometric redshifts provided by the \jades\ DR1 for all samples when spectroscopic redshifts are unavailable. We estimate an average offset of $<z_{\rm spec}-z_{\rm phot}>=0.03$ and $\sigma_{NMAD} = 0.024$ with about 13\% catastrophic outlier. Slightly higher outlier fraction for our companion sample is because of the confusion between Lyman break and Balmer break at  $z\sim3$.
	
	The results presented in this letter use a projected distance $d_{pro}<40\,kpc$ and velocity offset $|dv|<10000$\,km/s ($\sim 1.5\times\sigma_{NMAD}$ companions).   Figure \ref{fig:2d_maps} shows examples of companions identified around a subset of EELGs.  The primary conclusions of this paper do not change significantly if we vary the maximum $d_{pro}\ {\rm between}\ 20-50\,$kpc, and maximum $|dv|\ {\rm between}\ 5,000-20,000\,$km/s. Accuracy of photometric redshifts will strongly affect the properties of companions identified in this paper. About 75\% [61\%] of companions to EELGs [control] have $I>0.75$ where $I$ corresponds to the peakiness of the redshift probability distribution defined as $I = \int_{z_{\rm peak}-\Delta z}^{z_{\rm peak}+ \Delta z} P(z) dz$, where $\Delta z = 0.1*(1+z_{\rm peak})$. Our  main conclusions do not change significantly if we restrict the analysis to companions with $I>0.75$.

	Unfortunately, we do not have statistically significant number of spectroscopically confirmed EELGs or control galaxy pairs to test the companion identification technique (three true spectroscopic pairs out of four photometrically identified companions). We instead use cosmological simulations to quantify the accuracy of the companion identification technique (See Section \ref{sec:simulations}).

\begin{figure}
		\centering
		\tiny
		\includegraphics[scale=0.4, trim=0.0cm 0.0cm 0.0cm 0.0cm,clip=true]{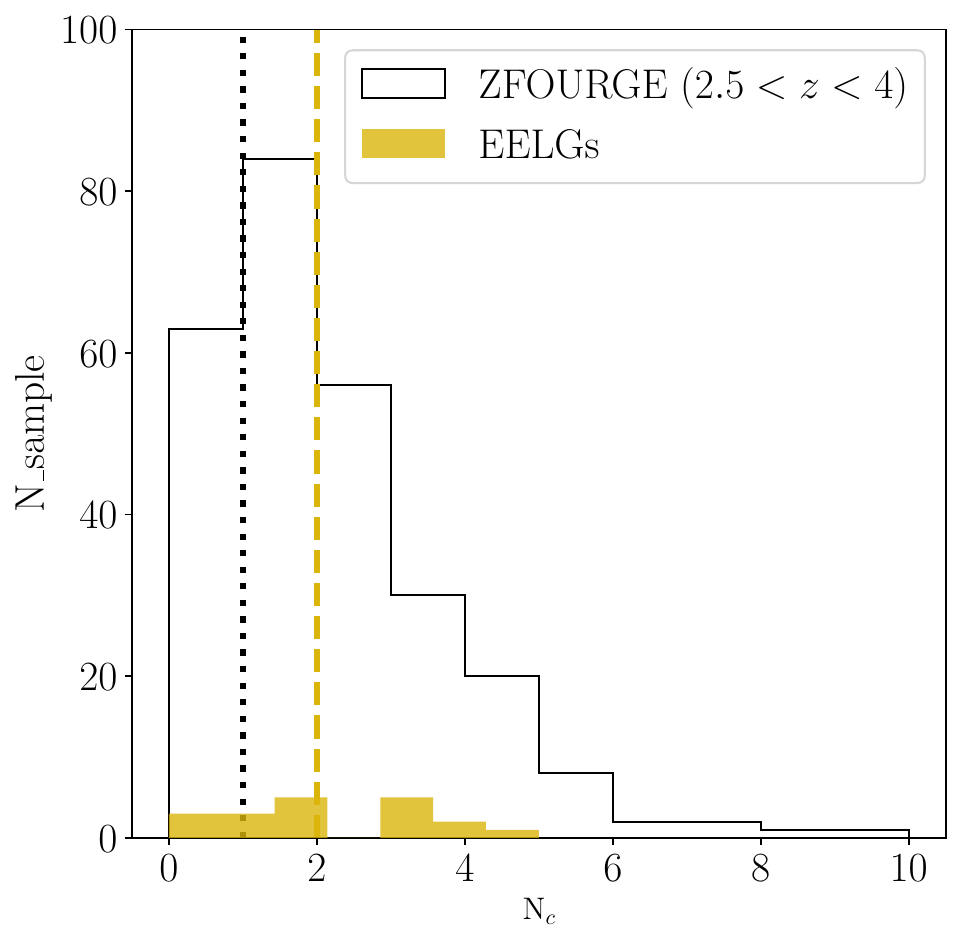}	
		
		\caption{Distribution of number of companions around EELGs (solid gold) and control sample (black). The dashed lines indicate the medians of the respective samples. EELGs on average have two companions whereas galaxies in the control sample only have one companion.}
		\label{fig:n_companions}
	\end{figure}

	The median number of companions around EELGs is two, whereas  galaxies in the control  sample only have one companion (Figure \ref{fig:n_companions}). About 16\% of the EELGs (3 of 19) do not have any companion as apposed to 24\%  in the control sample.  However, a standard two sided KS-test suggests no significant difference between the distribution of number of companions around EELGs and the control sample. For EELGs, we find a weak negative correlation between the sSFR and the nearest neighbour distance (Spearman's coefficient $\sim -0.6$, with $p-$value$=0.008$) but not for the control sample. We suspect that limited sample size and uncertainty in photometric redshift washes out this weak correlation.

	\begin{figure}
		\centering
		\tiny
		\includegraphics[scale=0.6, trim=0.0cm 0.0cm 0.0cm 0.0cm,clip=true]{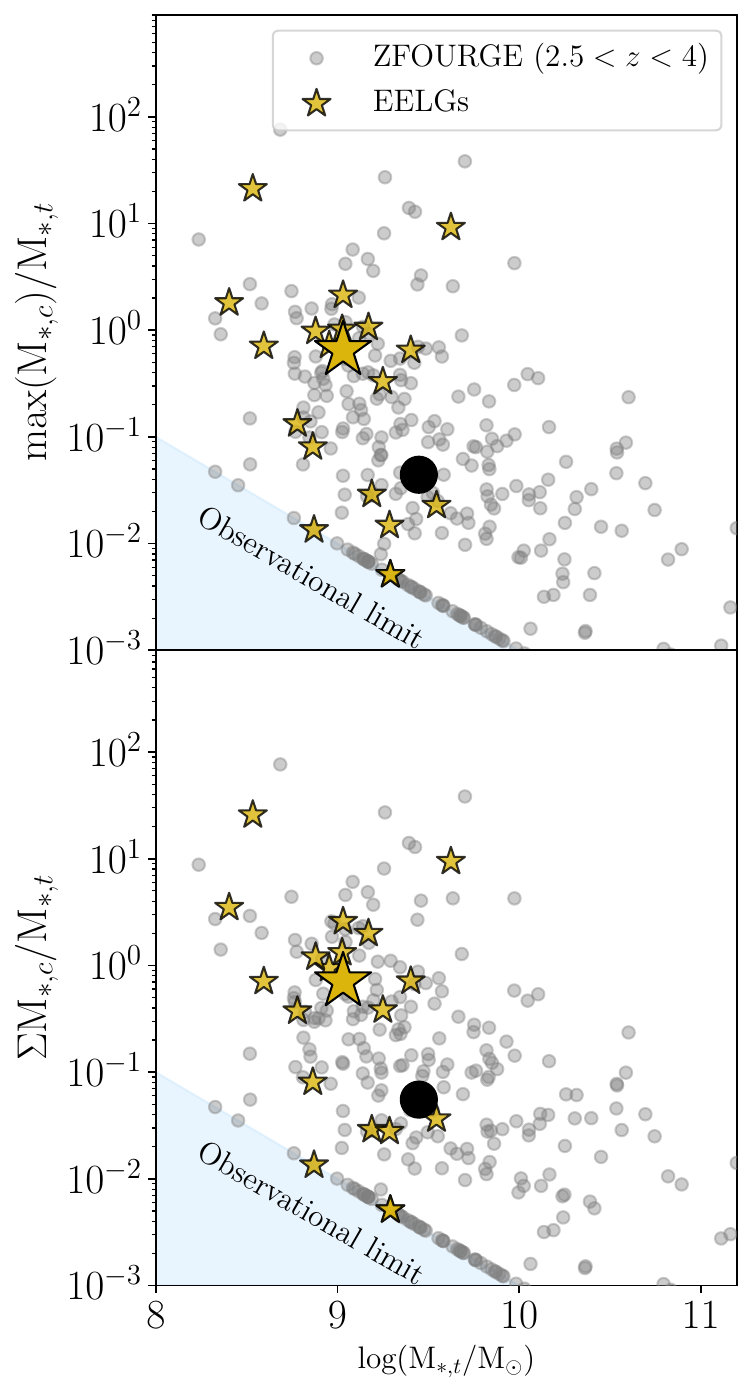}		
		\caption{The ratio of stellar mass of the brightest companion (top) and total stellar mass of all companions (bottom) to the target galaxy as a function of stellar mass of primary targets. Color scheme same as figure \ref{fig:companions_prop}. The larger symbols represent the median of the respective samples. The blue shaded region represent the stellar mass limit (\logmstar=7) of our companion population. A higher stellar mass ratio of companions around EELGs suggests EELGs might be undergoing strong interaction/major mergers.  }
		\label{fig:mass_ratio_companions}
	\end{figure}
	
	To determine whether the galaxies are experiencing strong or weak interactions, we calculate the ratio of the stellar mass of the most massive companion and the total stellar mass  of all companions to the target galaxy (Figure \ref{fig:mass_ratio_companions}). For galaxies without a detected companion, we assign a maximum companion mass of $10^7$\,\msun\ (minimum stellar mass estimated for the full companion population). The median stellar mass ratio of the most massive companion to EELGs is $\sim 0.65$, whereas it is only  $0.04$ for the control (two sided KS-test $p=0.007$) sample. The median stellar mass ratio of all companions combined is also less than $0.10$ for the control sample ($p=0.006$).

	We use bootstrapping to account for the differences in the sample size, stellar mass, and sSFR distribution of the EELGs and control sample. For each bootstrapped iteration and sample, we randomly select 19 galaxies (equal to the total number of EELGs) while allowing for repeats.   
	The stellar mass matching is done by randomly selecting one control galaxy for every EELG whose stellar mass is within 0.1\,dex of the EELG \citep{kaasinen_2016_COSMOSOIISurvey, gupta_2021_MOSELIllustrisTNGMassive}. An additional restriction on  sSFR to be within 0.1\,dex of the EELG is imposed for the stellar mass plus sSFR matched control sample. We estimate a KS-test $p>0.4$ between EELG and the respective property of the matched samples for all iterations.

\begin{figure*}
		\centering
		\tiny
		\includegraphics[scale=0.53, trim=0.0cm 0.0cm 0.0cm 0.0cm,clip=true]{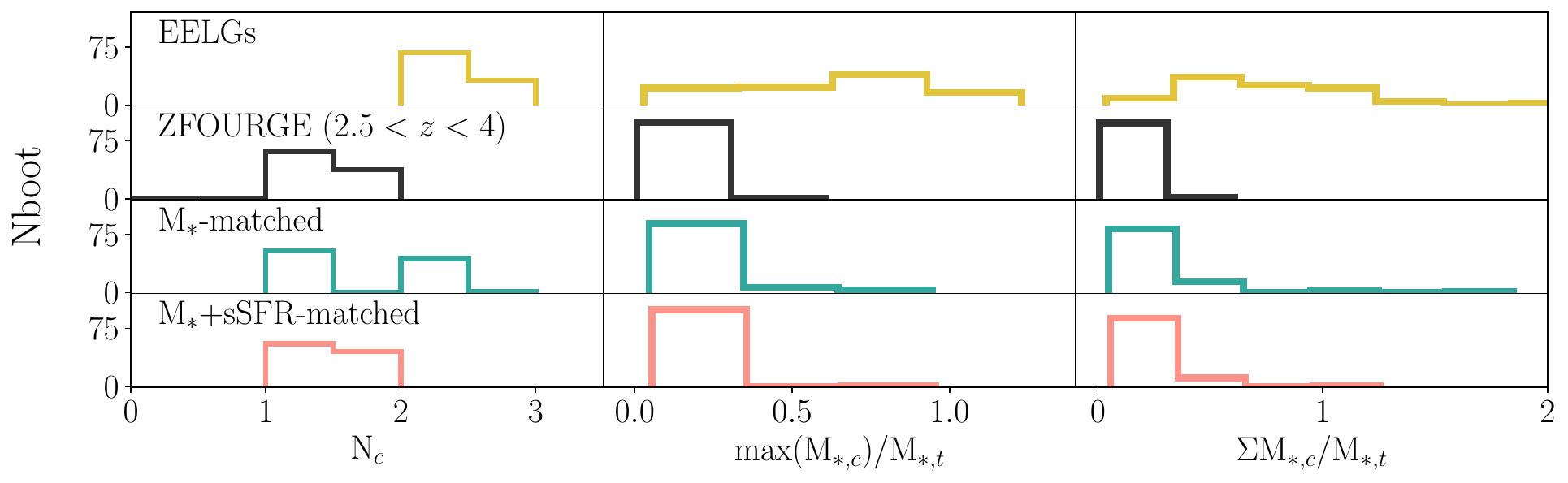}		
		\caption{Bootstrapped distribution of the median number of companions (left), the stellar mass ratio of the brightest companion (middle), and the total stellar mass ratio of all companions (right) for EELGs (top row), full control sample (second row), stellar mass matched (third row), and stellar mass plus sSFR matched samples (fourth row). EELGs have roughly similar stellar mass companion galaxies whereas whereas stellar mass ratio of companions for all control samples is 3-10 times smaller.} 
		\label{fig:mass_ratio_companions_boot}
	\end{figure*}

	On average EELGs have two companion galaxies and the stellar mass ratio of the most massive companion is 0.65 times the EELG (Figure \ref{fig:mass_ratio_companions_boot}). In contrast, the median mass ratio of the most massive companion and the total mass of all companions remains less than $0.34$ and $0.37$ respectively, for 90\% of the iterations (Figure \ref{fig:mass_ratio_companions_boot}) across all control samples. Even compared to the stellar mass plus sSFR matched sample, EELGs have three times more massive companions (Table \ref{tb:companion_properties}). 
	Thus, our measurements suggest that EELGs are more likely to be surrounded by similar stellar mass companions, and thus are more likely to be undergoing major mergers/ experiencing strong interactions.

	\subsection{False companion contamination}
	False projections could result in a higher fraction of companions being identified per galaxy. To estimate this effect, we randomly positioned both EELGs and control galaxies at 1000 different locations and re-identified companions at each location, following the criteria outlined in the previous section (projected distance $d_{\text{proj}} < 40$ kpc and relative velocity $|dv| < 10,000$ km/s). We detected at least one companion at approximately 30\% of the randomly chosen locations for both EELGs and control galaxies. The increased sensitivity of the \jades\ survey (4.5 nJy at $5\sigma$), leads to a significantly higher contamination fraction compared to previous studies  \citep{watson_2019_GalaxyMergerFractions}. 
	
	To determine the effect of contamination on the main conclusions of this paper, we generated 100 sets of the four samples (EELGs, control, mass-matched  and mass plus sSFR matched control sample) by randomly removing 30\% of the companions identified for each galaxy. For each iteration and sample, we calculated the median number of companions, the mass ratio of the brightest companion, and the total mass of all companions. We found very little overlap between the distributions of medians across all samples and parameters (KS-test $p < 0.001$).
	
	The median number of companions for EELGs remained one, whereas for the full control sample, after removing 30\% of companions, the median number became zero  in all iterations. Similarly for the stellar mass- and stellar mass plus sSFR-matched control samples, the median number of companions became zero in about 50\% of the iterations. The median mass ratio of the brightest companion [the median ratio of the total stellar mass of all companions] was $0.13_{-0.07}^{+0.01}$ [$0.22_{-0.16}^{+0.02}$], $0.03_{-0.01}^{+0.02}$ [$0.03_{-0.01}^{+0.03}$], $0.03_{-0.01}^{+0.03}$ [$0.03_{-0.01}^{+0.04}$] and $0.008_{-0.001}^{+0.001}$ [$0.008_{-0.001}^{+0}$] for EELGs, mass-matched, mass plus sSFR matched and full control samples, respectively. After removing 30\% of companions, EELGs still have almost three [six] times higher mass ratio of the brightest companion [total stellar mass of all companions]  than the stellar mass and stellar mass plus sSFR-matched control sample. Therefore, assuming a 30\% false companion contamination in our sample does not change the main conclusion of this paper.

    \begin{table}[t]
	\small
	\begin{center}
		\caption{Properties of companions}
		\label{tb:companion_properties}
		\begin{tabular}{ |p{2.cm}|  p{1.0cm}|   p{2.2cm}| p{1.8cm}|}
			\hline
			Sample & N$_c^{a,b}$ &  max(M$_{*,c}$)/M$_{*,t}^{a,c}$     & $\Sigma$M$_{*,c}$/M$_{*,t}^{a,d}$ \\
			\hline
			EELGS & $2_{-0}^{+1}$ &  $0.65_{-0.51}^{+0.11}$  &$0.71_{-0.34}^{+0.27}$ \\
			ZFOURGE ($2.5<z<4$) & $1_{-0}^{+1}$&  $0.05_{-0.02}^{+0.05}$ &$0.07_{-0.04}^{+0.04}$\\
			M$_*$-matched & $1_{-0}^{+1}$ &  $0.15_{-0.03}^{+0.12}$ & $0.20_{-0.08}^{+0.17}$ \\
			M$_*$-sSFR-matched & $1_{-0}^{+1}$ &  $0.17_{-0.07}^{+0.08}$ & $0.22_{-0.08}^{+0.12}$ \\
			\hline
		\end{tabular}
	\end{center}
	\begin{flushleft}
		{\bf Notes:}\\
		$^a$: Numbers represent the $50^{th},\ 16
		^{th},\ {\rm and}\ 84^{th}$ percentile errors.\\
		$^b$: N$_c$ is the number of companions.\\
		$^c$: max(M$_{*,c}$)/M$_{*,t}$ is the stellar mass of the most massive companion to the target galaxy.\\
		$^d$: $\Sigma$ M$_{*,c}$/M$_{*,t}$ is the total stellar mass of all companions to the target galaxy.   
	\end{flushleft}
\end{table}

	\subsection{Test with TNG100 simulations}\label{sec:simulations}
	To test the robustness of our companions identification technique, we use TNG100 simulation, which is part of the IllustrisTNG suit of cosmological simulations  \citep{springel_2018_FirstResultsIllustrisTNG, naiman_2018_FirstResultsIllustrisTNG, marinacci_2018_FirstResultsIllustrisTNG, pillepich_2018_FirstResultsIllustrisTNG, nelson_2018_FirstResultsIllustrisTNG}. From TNG100, we select all galaxies with stellar and dark matter mass $>10^8$\,\msun\ at snapshot 25, i.e., $z=3$. 
	
	We use the projected distance along $xy-$axis and velocity separation along $z$-axis to identify companions around each galaxy. At $d_{xy}
	<40\,{\rm kpc}, |dv_z|<10000\,$km/s 72\% of the galaxies have more than one companion, and for 61\% of the galaxies all identified companions live in the same halo as the target galaxy. If we restrict $d_{xy}$ to 20\,kpc then only 19\% of the galaxies have more than one companion, and for 85\% of galaxies all companions live in the same halo as the target galaxy. At $d_{xy}<$\,60\,kpc all galaxies have more than one companion, but only 29\% of galaxies all companions share the same halo as the target galaxy. The velocity separation only affects these measurements at $|dv_z|<200\ {\rm km/s}$ ($\sim 0.03\sigma_{NMAD}$ companion), which is not practical for observations given the uncertainty in photometric redshifts. 
	
	We use the merger history trees \citep{rodriguez-gomez_2015_MergerRateGalaxies} to estimate the fraction of  companions with same descendent as the target galaxy. At $d_{xy}=40\,{\rm kpc}, |dv_z|<10000\,$km/s, for about 63 percent of galaxies all identified companions will merge into a common descendent. Also, for 99.8\% of galaxies at least one companion has the common descendent as the target galaxy. Our measurements are similar to \cite{snyder_2017_MassiveClosePairs} who used a similar photometric redshift and projected distance approach on mock surveys with simulated data to find about 80\% of galaxy pairs identified at $z=2$ will merge by $z=0$. This suggest that the projected phase-space approach used in this work can successfully identify galaxies that eventually merge together and might already be  experiencing strong interactions.

	\section{Discussion and Conclusion}\label{sec:discussion}
	
	This paper utilizes deep JWST/NIRCam photometry and accurate photometric redshifts from the \jades\ survey to demonstrate that major mergers and/or strong interactions may be driving the extreme emission lines. We analyze the properties of companion galaxies (projected distance $< 40$\,kpc, velocity separation of $< 10,000$\,km/s) around 19 EELGs and \ncontrolJades\ control galaxies at redshifts $2.5 < z < 4$. Tests conducted using the TNG100 simulation confirm that nearly all galaxies will eventually merge with at least one companion galaxy by $z = 0$ (Section \ref{sec:simulations}), affirming the robustness of our companion identification technique.
	
	We find that the median mass ratio of the most massive companion and the total mass ratio of all companions around EELGs is $0.65_{-0.51}^{+0.11}$ and $0.71_{-0.34}^{+0.27}$ respectively. In contrast, for control galaxies at similar redshifts these ratios are only $0.05_{-0.02}^{+0.05}$ (KS test $p=0.007$) and $0.07_{-0.04}^{+0.04}$ ($p=0.006$).  Even after comparing  with a stellar mass and stellar mass plus sSFR-matched control sample, EELGs have three-five times higher mass ratios of the brightest companion and total mass of all companions (Figure \ref{fig:mass_ratio_companions_boot}). Our measurements suggest that EELGs are more likely to be surrounded by relatively more massive companion galaxies. We need spectroscopic data to confirm whether galaxies are undergoing major mergers or just experiencing strong interactions.
	
	Mergers and galaxy-galaxy interactions can induce the circumgalactic medium gas to cool down and boost the star formation rate by 30-40\% as shown by detailed hydro-dynamical simulations 
	\citep{moreno_2019_InteractingGalaxiesFIRE2, sparre_2022_GasFlowsGalaxy}.
	Some studies find evidence of bursty/rising star formation histories in EELGs \citep{cohn_2018_ZfourgeExtreme5007, endsley_2023_StarformingIonizingProperties}.  We find that companions around EELGs are relatively more massive even compared to the stellar mass plus sSFR matched sample. This suggest that extreme emission lines might be produced at significantly shorter timescale than the typical timescale of SFRs estimated from the SED models ($\sim$100\,Myrs). We suspect that gas cooling induced by strong interactions/mergers could be triggering the starburst episodes, which, in turn, produces the extreme emission lines.
	
	Cosmological simulations predict an almost two-order increase in the merger rate between  $z=0-6$ \citep{hopkins_2010_MergersBulgeFormation, rodriguez-gomez_2015_MergerRateGalaxies}. Deep photometric investigations have also confirmed the monotonic increase in the merger rate till $z=6$ \citep{duncan_2019_ObservationalConstraintsMergera}. We hypothesise that the increased merger rate might be responsible for the overabundance of EELGs detected with JWST at $z>6$ \citep{endsley_2023_StarformingIonizingProperties, cameron_2023_JADESProbingInterstellar, tang_2023_JWSTNIRSpecSpectroscopy, rinaldi_2023_MIDISStrongHv}. Thus, properly accounting for mergers would be important while estimating physical properties of gas and stars, in particular their kinematics and morphology in the early universe.

	\section{Acknowledgments}
	
	This research were supported by the Australian Research Council Centre of Excellence for All Sky Astrophysics in 3 Dimensions (ASTRO 3D), through project number CE170100013. TN acknowledge support from Australian Research Council Laureate Fellowship FL180100060. AH acknowledges support from the ERC Grant FIRSTLIGHT and Slovenian national research agency ARRS through grants N1-0238 and P1-0188. MH acknowledges funding from the Swiss National Science Foundation (SNF) via a PRIMA Grant PR00P2 193577 “From cosmic dawn to high noon: the role of black holes for young galaxies”

	\vspace{5mm}
	
	\appendix
	
{\bf     \begin{table}[t]
		\small
		\begin{center}
			
			\caption{EELGs and properties of their companions galaxies. }
			\label{tb:eelgs_properties}
			\begin{tabular}{ |p{2.cm}|  p{2.0cm}|   p{2.0cm}| p{1.5cm}|p{1.5cm}|   p{2.2cm}| p{2.0cm}|}
				\hline
				ZFOURGE ID & RA(J2000) & DEC(J2000) & z & N$_c$ &  max(M$_{*,c}$)/M$_{*,t}^{c}$     & $\Sigma$M$_{*,c}$/M$_{*,t}^{d}$ \\
				\hline
				12533 & 53.1426 & -27.8266 & 3.568$^a$ & 3.0 & 9.19 & 9.43 \\
				12903 & 53.1795 & -27.8239 & 3.17$^b$ & 4.0 & 21.2 & 25.7 \\
				17189 & 53.1983 & -27.7892 & 3.55$^a$ & 3.0 & 1.81 & 3.49 \\
				17342 & 53.1434 & -27.7881 & 3.41$^b$ & 0.0 & 0.0135 & 0.0135 \\
				18742 & 53.1596 & -27.7768 & 3.436$^a$ & 4.0 & 0.133 & 0.374 \\
				11398 & 53.1533 & -27.836 & 3.56$^b$ & 1.0 & 0.708 & 0.708 \\
				11548 & 53.165 & -27.8339 & 3.11$^b$ & 2.0 & 0.722 & 0.939 \\
				15357 & 53.1408 & -27.8041 & 2.616$^a$ & 0.0 & 0.00512 & 0.00512 \\
				17583 & 53.1808 & -27.7863 & 2.69$^a$ & 2.0 & 1.02 & 1.32 \\
				18053 & 53.1958 & -27.7828 & 3.326$^a$ & 1.0 & 0.0806 & 0.0806 \\
				19656 & 53.2034 & -27.7704 & 2.71$^b$ & 2.0 & 0.649 & 0.714 \\
				19863 & 53.17 & -27.7684 & 3.087$^a$ & 1.0 & 0.029 & 0.029 \\
				20257 & 53.165 & -27.7652 & 3.192$^a$ & 2.0 & 0.0228 & 0.0364 \\
				22136 & 53.1529 & -27.7493 & 3.088$^a$ & 5.0 & 1.06 & 2.0 \\
				22277 & 53.1493 & -27.7487 & 2.524$^a$ & 0.0 & 0.00511 & 0.00511 \\
				13155 & 53.1757 & -27.8223 & 3.064$^a$ & 3.0 & 0.977 & 1.2 \\
				13203 & 53.1535 & -27.8215 & 3.563$^a$ & 3.0 & 2.13 & 2.58 \\
				15111 & 53.1612 & -27.8062 & 2.987$^a$ & 2.0 & 0.0148 & 0.0278 \\
				16603 & 53.1313 & -27.793 & 3.61$^b$ & 3.0 & 0.33 & 0.382 \\        \hline
			\end{tabular}
		\end{center}
		\begin{flushleft}
			{\bf Notes:}\\
			$^a$: Spectroscopic redshift from the \jades\ DR1 or \mosel\ survey\\
			$^b$: Photometric redshift from the \jades\ DR1.\\
			$^c$: max(M$_{*,c}$)/M$_{*,t}$ is the stellar mass of the most massive companion to the target galaxy.\\
			$^d$: $\Sigma$ M$_{*,c}$/M$_{*,t}$ is the total stellar mass of all companions to the target galaxy.   
		\end{flushleft}
	\end{table}
}

\end{document}